\documentclass[lettersize,journal]{IEEEtran}
\usepackage{amsmath,amsfonts}
\usepackage{algorithmic}
\usepackage{algorithm}
\usepackage{array}
\usepackage{multirow}
\usepackage[caption=false,font=normalsize,labelfont=sf,textfont=sf]{subfig}
\usepackage[pagebackref=false,breaklinks=true,letterpaper=true,colorlinks,bookmarks=false, linkcolor=red, filecolor=black, urlcolor=blue, citecolor=green]{hyperref}
\usepackage{textcomp}
\usepackage{stfloats}
\usepackage{url}
\usepackage[table,xcdraw]{xcolor}
\usepackage{verbatim}
\usepackage{graphicx}
\usepackage{cite}
\usepackage{graphicx}
\usepackage{amsmath}
\usepackage{amssymb}
\usepackage{booktabs}
\usepackage{comment}
\usepackage{multirow}
\usepackage{mathtools}
\usepackage{epsfig}
\usepackage{multirow}
\usepackage{tikz}
\usepackage{comment}
\usepackage{amsmath,amssymb} 
\usepackage{color}
\usepackage{bbding}
\usepackage{wrapfig}
\usepackage{threeparttable}
\newcommand{\etal}{\textit{et al.}}

\newcommand{\eg}{\emph{e.g., }}
\newcommand{\etc}{\emph{etc. }}
\usepackage[top=0.7in,bottom=0.5in,left=0.5in,textwidth=7.5in]{geometry}

\begin{document}

\title{\textcolor{black}{Boosting Binary Neural Networks via Dynamic Thresholds Learning}}

\author{Jiehua Zhang, Xueyang Zhang, Zhuo Su, Zitong Yu, Yanghe Feng, Xin Lu, Matti Pietik\"{a}inen, Li~Liu
\thanks{This work was partially supported by the National Key Research and Development Program of China No. 2021YFB3100800, the Academy of Finland under grant 331883 
and the National Natural Science Foundation of China under Grant 61872379 and 62022091. The CSC IT Center for Science, Finland, is also acknowledged for computational resources.}
\thanks{Jiehua Zhang and Xueyang Zhang have equal contributions. Li Liu (li.liu@oulu.fi) is the corresponding author.}
\thanks{Jiehua Zhang, Zhuo Su, and Matti Pietik\"{a}inen (\{ jiehua.zhang, zhuo.su, matti.pietikainen\}@oulu.fi) are with the Center for Machine Vision and Signal Analysis (CMVS) at the University of Oulu, Finland.}
\thanks{Xueyang Zhang, Yanghe Feng (\{zhangxueyang, fengyanghe\}@nudt.edu.cn), and Xin Lu (xin\_lyu@sina.com) are with the College of System Engineering, National University of Defense Technology (NUDT), China. Zitong Yu (zitong.yu@ntu.edu.sg) is with the ROSE Lab, Nanyang Technological University, Singapore. Li Liu is with the Laboratory for big data and decision at the College of System Engineering, NUDT, China and is also with Center for Machine Vision and Signal analysis at the University of Oulu, Finland.}}

\markboth{Under Review}
{Shell \MakeLowercase{\textit{et al.}}: A Sample Article Using IEEEtran.cls for IEEE Journals}


\maketitle

\begin{abstract}
\textcolor{black}{Developing lightweight Deep Convolutional Neural Networks (DCNNs) and Vision Transformers (ViTs) has become one of the focuses in vision research since the low computational cost is essential for deploying vision models on edge devices. Recently, researchers have explored highly computational efficient Binary Neural Networks (BNNs) by binarizing weights and activations of Full-precision Neural Networks.} \textcolor{black}{However, the binarization process leads to an enormous accuracy gap between BNN and its full-precision version. One of the primary reasons is that the Sign function with predefined or learned static thresholds limits the representation capacity of binarized architectures since single-threshold binarization fails to utilize activation distributions. To overcome this issue, we introduce the statistics of channel information into explicit thresholds learning for the Sign Function dubbed DySign to generate various thresholds based on input distribution}. Our DySign is a straightforward method to reduce information loss and boost the representative capacity of BNNs, which can be flexibly applied to both DCNNs and ViTs (\emph{i.e.}, DyBCNN and DyBinaryCCT) to achieve promising performance improvement. As shown in our extensive experiments. \textcolor{black}{For DCNNs, DyBCNNs based on two backbones (MobileNetV1 and ResNet18) achieve 71.2$\%$ and 67.4$\%$ top1-accuracy on ImageNet dataset, outperforming baselines by a large margin (\emph{i.e.}, 1.8$\%$ and 1.5$\%$ respectively). For ViTs, DyBinaryCCT presents the superiority of the convolutional embedding layer in fully binarized ViTs and achieves 56.1\% on the ImageNet dataset, which is nearly 9\% higher than the baseline.}

\end{abstract}

\begin{IEEEkeywords}
 Binary neural network, Vision transformer, Model compression, Object recognition
\end{IEEEkeywords}

\section{Introduction}
\textcolor{black}{In the past decade, Deep Neural networks (DNNs), especially earlier Deep Convolutional Neural Networks (DCNNs) \cite{he2016deep,simonyan2014very, krizhevsky2017imagenet,li2022deep} and more recent Vision Transformer (ViT) \cite{vit2021,survey2021}, have brought tremendous progress in numerous computer vision tasks including image classification \cite{vit2021,efficient2021,he2016deep,hu2018squeeze,li2021survey}, object detection  \cite{detr2020,liu2020deep,swin2021}, image segmentation \cite{segformer2021,SEtrans2021,minaee2021image}, and video classification \cite{super2021,phys2021,jin2020deep}, \etc, and are still developing dizzyingly fast. Progress in vision, however substantial, has been achieved by increasingly large
and complex DNNs \cite{vit2021,yu2022coca} that are computationally expensive, environmentally unfriendly (due to the massive carbon footprints), and cannot be deployed in resource-constrained edge devices like the Internet of Things, and portable devices. Therefore, to mitigate these challenges, efficient processing of DNNs \cite{8253600, deng2020model} has attracted enormous attention, and numerous methods have been proposed including model quantization \cite{fqvit2021,post2021,ternary2020}, network pruning \cite{prune2021,vtp2021,jiang2022model}, model distillation \cite{dis2021,efficient2021}, compact networks \cite{su2021pixel,parcnet2022} and efficient architecture design \cite{mobile2021,syn2021,aft2021,res2021}. }


\textcolor{black}{Among the aforementioned mainstreams for DNN compression, model
quantization \cite{gholami2021survey} aims at reducing the number of bits per weight and (or) activation for saving memory storage and computation via various forms of data quantization. As an extreme case of model quantization, Binary Neural Networks (BNNs) \cite{bnn2016,xnor2016,birealnet2018} binarize model weights and activations, which can drastically save computing resources and lead to considerable speedup. Therefore, BNNs have appeared as one of the most prominent solutions for implementing deep models on edge devices. Despite the great computational efficiency, the binarization results in severe information loss and more difficulty during model optimization. These lead to constrained representational capacity of BNNs and big accuracy drops in
comparison with their real-valued versions. Hence, how to train highly accurate BNNs remain open. If successful, accurate BNNs will enable ubiquitous vision applications at the edge near the sensor.}


\textcolor{black}{Existing studies on how to optimize BNNs can be grouped into the following categories \cite{qin2020binary}:
\begin{itemize}
  \item \textbf{Minimize the binarization error.} A common practice to minimize the quantization error for binarized weights and activations is to approximate full-precision parameters as closely as possible since directly using 1-bit (-1/+1) severely impairs the performance of BNNs. Existing researches \cite{xnor2016,zhou2016dorefa,wang2021bi, ding2019regularizing, reactnet2020} apply a scale factor for binary weights and regularize activation distribution. These methods can benefit the training of BNNs and reduce information loss in each layer to achieve a significant performance improvement.
  \item \textbf{Improve loss function.} Minimizing quantization error focuses on optimizing BNNs in each layer and omits the impact of the final output. Some works \cite{hou2016loss, shang2022network} regard binarization as a training objective and design a specific loss function associated with binary weights and activations to guide the learning of BNNs globally.    
  \item \textbf{Reduce the gradient error.} Existing BNNs are trained with the straight-through estimator (STE) technique to deal with the gradients for the Sign function. This method has two problems: 1) gradient mismatch between Sign and STE; 2) parameters not in [-1, +1] are unable to be updated. These problems seriously affect the optimization of BNNs. To solve these problems, some works \cite{qin2020forward, xu2021learning, yin2019blended} attempt to modify the binarization function in backward or forward propagation or calibrates the gradient to reduce the gradient error.    
\end{itemize}
}

\textcolor{black}{Despite the considerable research efforts, the limitations of existing research on BNNs include:
\begin{itemize}
  \item ignoring the gap in the distribution of diverse samples.
  \item focusing on binarizing DCNNs (BCNNs) rather than ViTs, GANs, \etc
  \item focusing on image classification rather than other vision tasks like object detection, semantic segmentation, \etc
\end{itemize}
}

\begin{figure*}[]
    \centering
    \includegraphics[width=\linewidth]{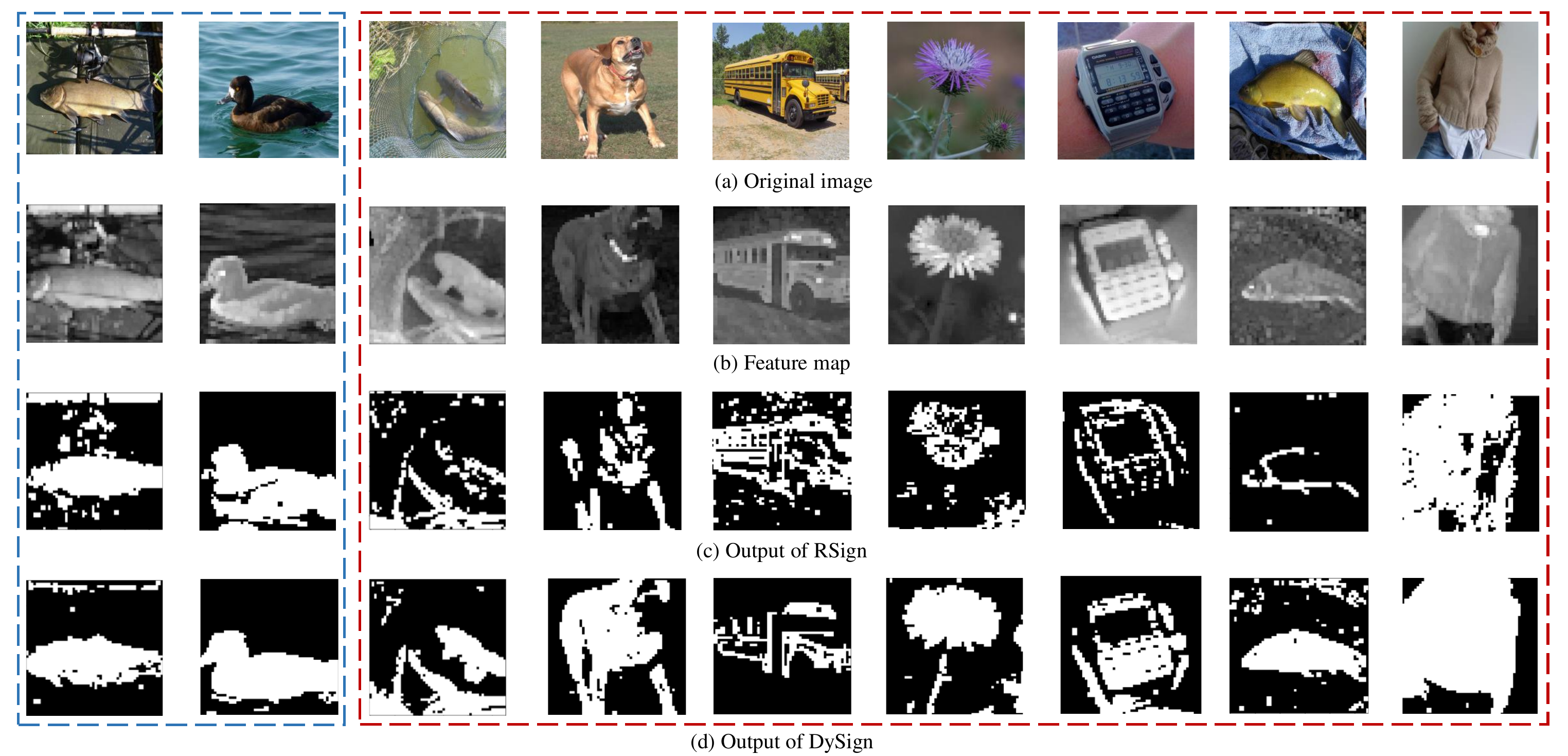}
    \caption{\textcolor{black}{The output of different samples after RSign and DySign. The samples (a) are randomly selected from the ImageNet validation dataset. (b) denotes the feature maps of sample (a) before passing through Sign Function, which is extracted from the shallow layer in binarized ResNet18. (c) shows the output of RSign. The activations of samples in the blue box can remain useful object information. However, RSign drops valuable object information and introduces excess background noise features for samples in the red box. In (d), the output of DySign can retain the feature information of the object since it incorporates adaptive thresholds to process features from diverse samples, which benefits training BNNs.}}
    \label{fig:Fig1}
\end{figure*}

\textcolor{black}{In this paper, we focus on improving the first two limitations. Many existing BNNs apply the same threshold value (like 0) to binarize all feature maps of each input image. But this might not be appropriate in all cases, \eg activation feature maps highlight different information of the input image, and have different feature distributions. This one-for-all (one threshold for all feature maps and all images) binarization has serious limitations such as failing to adapt to feature map distributions.
It is easy to realize \cite{reactnet2020} that the binarized feature map is sensitive to the threshold value used for binarization, as the fluctuation above or below the threshold can cause the binarized feature map to have a different appearance. In this case, adaptive thresholding is beneficial. Therefore, Liu \etal \cite{reactnet2020} proposed ReActNet that learns a threshold for each feature map channel by a generalization of Sign function (RSign) and produces higher accuracy. }However, the learnable thresholds in the RSign are image-agnostic and still neglect activation variations of diverse samples, which makes ReActNet only adapt to feature channels, not input images. For various input images, the image-agnostic threshold fails to preserve representative features since considerable variations in their feature distributions. As shown in Fig. \ref{fig:Fig1}, the thresholds in RSign fit with images in the first two columns but drop valuable object information and introduce excess background noise features for other images, which limits representation learning of BNNs.

\textcolor{black}{To address this issue, the learning of thresholds should be based on statistics of channel information in each layer, which is equivalent to learning sample-independent thresholds. The sample-independent thresholds can shrink the gap between binarized and full-precision features by reshaping activation distributions based on input. To achieve this goal, we take the global channel information as a statistical criterion for generating sample-independent thresholds. A lightweight computational unit is incorporated with the Sign function to assign thresholds for each channel based on the statistical criterion. We name this novel Sign function as DySign, which can adaptively perform activation distributions adjustment.} Our DySign has two advantages: 1) generates diverse thresholds automatically according to feature distributions without human design; 2) can be flexibly transferred to other BNN architectures. Compared with RSign, the outputs of DySign remain informative features and suppress noise as shown in the last row of Fig. \ref{fig:Fig1}.

Existing BNNs in vision tasks are Binary CNNs (BCNNs). ViTs perform remarkable progress in vision tasks. However, this quadratic complexity limits ViTs to model high spatial-resolution images and deploy them on portable devices, which is crucial for vision tasks. Thus, we attempt to build a fully binarized ViT to reduce huge computational costs. We first observe that  replacing the linear layer in the patch embedding layer with a convolutional block significantly improves the performance of binary ViTs since local information can be well preserved. Therefore, we then construct a binarized ViT baseline based on Convolutional Transformer \cite{cct2021,coat2021,cmt2021,xiao2021early}. However, the binarization error in features seriously affects the self-attention module to capture the relationship between tokens. Thus, we adopt the DySign on both the self-attention layer and the feed-forward layer in each binarized transformer block to emphasize valuable information in each token.

This paper presents a substantial extension of our previous short conference paper \cite{dybnn2022} in ICASSP 2022\footnote{Our ICASSP version has only four pages excluding references. Therefore, this paper is indeed a substantial extension of our short conference version.}. In addition to the contribution Dynamic Binary Neural Network (DyBCNN) to generate diverse thresholds based on the activation distribution of the input itself in our conference paper, this paper presents the following additional technique contributions: 1) we conduct a first study to develop binary ViTs by proposing BinaryCCT building upon Compact Convolutional Transformer (CCT) \cite{cct2021} which is a fully binarized transformer architecture; 2) we analyze the impact of binarization for standard ViTs and Convolutional Transformers; 3) we present DyBinaryCCT by introducing the DySign to input tokens in BinaryCCT to improve the learning ability of BinaryCCT. 


The main contributions can be summarized as follows: 

\begin{itemize}
      \item We pioneer to propose a fully binarized transformer-based architecture for computer vision tasks. Compared with standard ViT, inserting a convolution based patching method in the tokenization layer can preserve local information between image patches and mitigate information loss caused by binarization. Thus, we build the baseline based on CCT, named BinaryCCT. 
     \item Based on the standard Sign function, we propose Dynamic Sign (DySign), a simple and effective method to enhance BNNs performance. DySign aggregates global activation distribution to produce an embedding, allowing information from the global receptive field of feature maps to be used to adaptively assign the thresholds for each channel. In addition to the performance advantages, DySign can be flexibly migrated to other BNN frameworks. With BCNN and BinaryCCT,  we propose DyBCNN and DyBinaryCCT, replacing the original Sign with DySign. The results suggest the superiority of our method.  
    \item Extensive experimental evaluations on the commonly used datasets for image classification. For DyBCNN, we demonstrate the effectiveness of on the ImageNet dataset. The DyBCNN is built on two networks (ReActNet and ReActNet based on ResNet18), achieving $71.2\%$ and $67.4\%$ top1-accuracy, outperforming strong baselines $1.8\%$ and $1.5\%$ by a large margin, respectively. For DyBinaryCCT, we evaluated on CIFAR10, CIFAR100, and ImageNet dataset. The DyBinaryCCT can achieve $84.98\%$  on CIFAR10, $62.10\%$ on CIFAR100, and 56.09\% on ImageNet, outperforming average $7\%$ over baseline. The proposed DySign significantly advances the performance of BNNs and flexibly migrate to various BNN frameworks.
\end{itemize}



The remainder of this paper is organized as follows. Section \ref{Sec:2} reviews the related works of binary neural networks and vision transformer. Section \ref{Sec:3} formulates the proposed DySign, and the architectures of DyBCNN and DyBinaryCCT. Section \ref{Sec:4} provides rigorous ablation studies and evaluates the performance of the proposed DyBCNN and DyBinaryCCT on the image classification task, respectively. Finally, the conclusion and future work are presented in Section \ref{Sec:5}.

\begin{figure*}[ht]
\centering
\includegraphics[width=18cm]{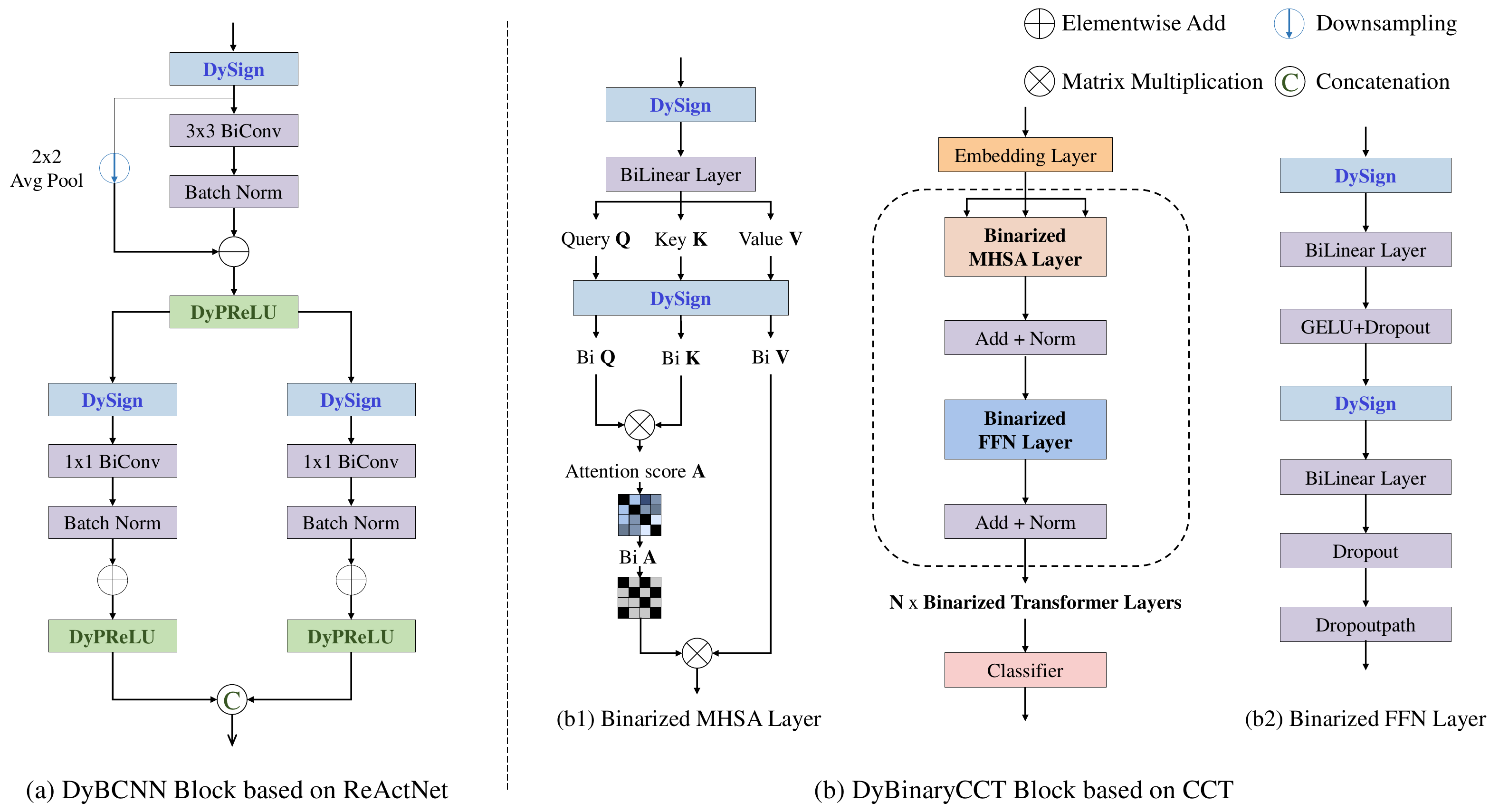}
\caption{\textcolor{black}{\textbf{(a)} denotes the basic module of proposed DyBCNN (built on the ReActNet). The DySign and DyPReLU replace original RSign and RPReLU; \textbf{(b)} denotes the components of proposed DyBinaryCCT based on CCT, consisting of three parts: 1) Embedding layer: a convolutional block to generate tokens; 2) Binarized Transformer layers: (b1) binarized Multi-Head Self-Attention (MHSA) layer and (b2) binarized Feed-Forward Network (FFN) layer; 3) Classifier to output logits.}}
\label{fig:fig2}
\end{figure*}

\section{Related Work}
\label{Sec:2}
\subsection{Binary Neural Networks}
The binary neural networks (BNN) serves as a favorable technique for deploying deep learning models on portable devices. Soudry \emph{et al.} \cite{EBP2014} claimed that good performance could be achieved when all neurons and weights are binarized. They proposed the expectation backpropagation (EBP) method to train the network with discrete values. Hubara \emph{et al.} \cite{bnn2016} binarized the weights and activations in networks and illustrated its success on the small datasets. While BNN largely saves storage and computation, the accuracy gap between BNN and full-precision model limits its application, especially when evaluated on large-scale dataset, like ImageNet. This issue is caused by the severe information loss and difficulty to the optimization. 

In recent years, lots of works attempted to improve the accuracy by introducing extra feature information to reduce quantization error and information loss. Rastegari \emph{et al.} \cite{xnor2016} proposed XNOR-Net, which incorporated scaling factors to multiply with binary kernel weights to reduce binarized error. Based on this, Liu \emph{et al.} \cite{birealnet2018} introduced an extra shortcut operation between continuous binary convolutional blocks to enhance representational capacity and reduce information loss of BNNs. Liu \emph{et al.} \cite{reactnet2020} also demonstrated that BNNs were sensitive to activation shift. They proposed simple activation distribution shift functions and built BNN based on MobileNetV1, resulting in a significant accuracy improvement. Martinez \emph{et al.} \cite{real2020} introduced a learning branch to compute the scale factors to rescale the activations. Besides, several methods also attempt to alleviate the optimization difficulty by using approximate gradient. IR-Net \cite{qin2020forward} enhances the performance of BNNs by retaining both forward and backward information. Xu \emph{et al.} \cite{xu2021learning} estimated the gradient of sign function in the frequency domain and proposed frequency domain approximation (FDA), which can better maintain the main direction of factual gradient. We considered to improve representational power of BNNs by replacing previous fixed and static thresholds with dynamic thresholds.
\subsection{Vision Transformer}
Transformer \cite{transformer2017} is firstly proposed to model sequential data for the machine translation task. Recently, the Vision Transformer (ViT) \cite{vit2021} is proposed by splitting an image into patches and providing the sequence of linear embedding of these patches as an input to the transformer-based architecture. Many other ViT variants \cite{cross2021,tit2021,dis2021,survey2021,swin2021,py2021,token2021} are then proposed and achieve great performance. The challenges of ViT involve: 1) The ViT is data-hungry and fails to generalize well on small datasets due to the lack of inductive bias of CNN, which can help model effectively process image; 2) The ViT requires huge memory and computation, which is intolerable for edge devices. 

To address the first problem, recent efforts attempt to introduce the CNN inductive bias to enhance Transformer. Touvron \emph{et al.} \cite{efficient2021} proposed a Data-efficient Transformer (DeiT) to free ViT from dependency on large-scale training data. Except for existing training strategies for data augmentation and regularization, a distillation strategy is applied for adjective representation learning.  Researchers also combine the transformer with CNN \cite{coat2021,cmt2021,cct2021,xiao2021early,res2021,d2021convit,chu2021conditional,yuan2021incorporating,li2021localvit} to leverage the strength of both CNNs and Transformers. These methods replace the linear patch project with the convolution stacks. Extensive experiments illustrate that the locality and translation invariance of CNN can effectively boost the performance of ViT across multiple datasets. Besides, a local attention mechanism is proposed to augment the local image information extraction ability \cite{swin2021,deform2021,tit2021,yuan2021volo,zhang2021multi}. One of the representative methods is the Swin Transformer \cite{swin2021}, which conducts a shifted window to model the global and local features in the spatial dimension. The shifted windows bridge the windows of the preceding layer, providing connections among them that significantly enhance modeling power. TNT proposed by Han \emph{et al.} \cite{tit2021} aggregates patch-level and pixel-level representation by two blocks: an inner block to capture interaction within each patch and an outer block to extract global information.  

For the second problem, several works have focused on lightweight model design \cite{mehta2021mobilevit,ma2022mocovit,yang2022lite,huang2022lightvit} and quantization \cite{post2021,fqvit2021}. MobileViT \cite{mehta2021mobilevit} attempts to incorporate attention into MobileNetV2 to improve mobile CNNs. presents a different perspective on the global processing of information with transformers. Huang \emph{et al.} \cite{huang2022lightvit} proposed the LightViT based on a pure transformer without convolution. For quantization, Liu \emph{et al.} \cite{post2021} adopted the post-training quantization algorithm to compress transformer architectures into 8-bit to reduce storage memory and computational cost. Lin \emph{et al.} \cite{fqvit2021} proposed FQ-ViT to achieve 4-bit ViT with comparable accuracy degradation on fully quantized ViT. In the NLP field, researchers have compressed the transformer model to 1-bit. Bai \emph{et al.} \cite{bi2020} compressed the weight and embedding of BERT to 1-bit. Qin \emph{et al.} \cite{bibert2022} and Liu \cite{liu2022bit} proposed fully binarized BERT and provided theoretical formulation and empirical observation for the degradation of the information of attention weight. In this paper, we focus on binarizing the vision transformer model for the visual recognition task.   

\section{Proposed Method}
\label{Sec:3}
In this section, we will first introduce the basic knowledge of binary neural networks and Compact Convolutional Transformer (CCT). Then, we present detail information about the proposed DySign. Based on this, then we introduce architectures of DyBCNN and DyBinaryCCT, which can be observed in Fig. \ref{fig:fig2}. The detailed approach for each aspect will be introduced following.   

\subsection{Preliminary}

\textbf{Binary Neural Networks.} BNNs utilize the Sign function to constrain weights and activations to +1 or -1. For explaining the operation of binarized convolution, we denote  $W^\ell$ and $X^\ell$ as the weights and input features in the \textsl{$\ell$}-layer. The input of the \textsl{$\ell+1$}-layer can be defined as:
\begin{equation}
    X^{\ell+1} =\phi^\ell(\text{Sign}(W^\ell) \otimes \text{Sign}(X^\ell)),\\
\end{equation}
\begin{equation}
   \text{Sign}(x)=\left\{
    \begin{aligned}
+1 & , & x>0, \\
-1 & , & x\le 0,
\end{aligned}
\right.
\end{equation}
where $\otimes$ denotes convolutional operation, the $\phi^\ell(\cdot)$ denotes the nonlinear operation in the \textsl{$\ell$}-layer (ReLU, PReLU, and BN layer, etc.). To enhance the performance of BNNs, the ReActNet \cite{reactnet2020} utilized the learnable channel-wise thresholds to shift feature maps for retaining valuable information, which can be expressed as below: 

\begin{equation}
   \text{RSign}(x_i)=\left\{
    \begin{aligned}
+1 & , & x_{i}>a_{i}, \\
-1 & , & x_{i}\le a_{i},
\end{aligned}
\right.
\end{equation}
where $x_{i}$ denotes an element of $i$-th channel in input feature map, $a_{i}$ denotes the threshold of  $i$-th channel, which illustrates that the threshold can vary for different channels. The PReLU function is also processed by this operation to reshape the feature map, which can be expressed as:

\begin{equation}
   \text{RPReLU}(x_i)=\left\{
    \begin{aligned}
x_{i}-\gamma_{i}+\zeta_{i} & , & x_{i}>\gamma_{i}, \\
\beta_{i}(x_{i}-\gamma_{i})+\zeta_{i} & , & x_{i}\le \gamma_{i}.
\end{aligned}
\right.
\end{equation}
where $\gamma_{i}$ and $\zeta_{i}$ denote the learnable shift parameters for reshaping the distribution. By introducing the activation distribution shift and reshape, ReActNet achieves a significant accuracy increase in image classification.

\vspace{0.4em}
\noindent \textbf{Compact convolutional transformer.} Vision Transformer (ViT) \cite{vit2021} is the first work to construct a pure transformer architecture for computer vision tasks. ViT consists of several components: Image Tokenization, Positional Embedding, Classification Token, the Transformer Encoder, and a Classification Head. The motivation of ViT is that it takes a sequence of image patches as the input and utilizes the self-attention block to capture the long-range relationship. However, ViT is only applicable when trained on a large-scale dataset, and it generalizes poorly when trained on insufficient amounts of data due to the lack of inductive biases. 

To reduce the dependency upon data, the Compact Convolutional Transformer (CCT) \cite{cct2021} introduces a convolutional based patching method and achieves prominent performance on the small dataset. CCT is composed of several parts: Convolution Block, Transformer Encoder, Sequence Pooling, and a Classification Head. The position embedding is optional in CCT. Detailed information on these components is presented below:

\vspace{0.4em}
\subsubsection{Convolution block} To improve the generalization ability of vision transformers on the limited training data, CCT replaces the image patching and projection layers in the ViT with a simple convolutional block. The convolutional block consists of a single convolutional layer, ReLU activation, and a max pooling layer. Compared with ViT directly patching on images, the convolutions can embed the image into a latent representation, which is more efficient. Furthermore, the convolution block can inject inductive bias into the model and preserve local spatial information. With these characteristics, CCT can provide competitive results on a small dataset.

\vspace{0.4em}
\subsubsection{Transformer encoder} The transformer encoder layer remains the same as the original Transformer \cite{transformer2017} and ViT \cite{vit2021}, consisting of a Multi-head attention (MHSA) layer and a Feed-Forward (FFN) layer. It utilizes Layer Normalization \cite{ln2016}, GELU activation, and dropout. The positional embedding is defaulted as learnable. 

\vspace{0.4em}
\subsubsection{Sequence pooling} Sequence Pooling (SeqPool) aims at pooling the sequential based information from transformer encoder layers and can eliminate the need for the classification token \cite{cct2021}. SeqPool can be regarded as the mapping transformation: $\mathbb{R}^{b\times n\times d}\rightarrow \mathbb{R}^{b\times d}$. It makes CCT weigh the sequential embeddings of the latent space from the transformer encoder and capture the relationship between the input data.

\begin{figure*}[!ht]
\centering
\includegraphics[width=0.8\linewidth]{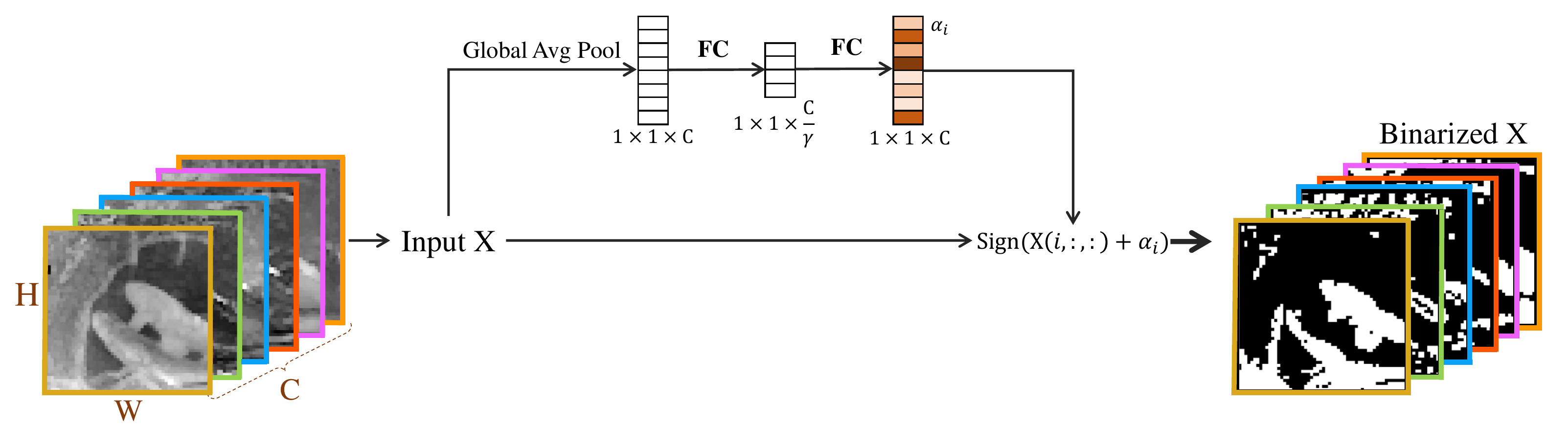}
\caption{\textcolor{black}{The DySign in the DyBCNN. DySign is a lightweight computational unit. The input feature map \textbf{X} first passes through the global pooling layer. Then, adaptive thresholds $\alpha_{i}$ are generated based on global information of \textbf{X} for each channel to shift activation distributions. Finally, the Sign function is applied to binarize shifted \textbf{X}. The computational unit consists of one global average pooling layer and two linear layers.}}
\label{fig:fig3}
\end{figure*}

\subsection{DySign: Dynamic Sign Function}

\textcolor{black}{Although previous works have achieved remarkable progress in BNNs, the feature information loss still severely degrades representation learning of BNNs. \textcolor{black}{The ReActNet focuses on activation distribution and adopts static channel-wise thresholds on the input feature maps of each layer. However, these image-agnostic thresholds neglect the difference between various images and fail to remain informative features for all cases. As observed Fig. \ref{fig:Fig1}, the RSign can preserve object information for the samples in the blue box but still eliminate informative features for the samples in the red box. To address this issue, BNNs requires statistical channel information to adjust thresholds adaptively when processing different input.} This motivates us to insert a computational unit in the Sign function to calculate thresholds based on the activation distribution of input feature maps. We name this novel Sign function as DySign.}

\textcolor{black}{We first introduce the DySign in BCNNs. The DySign can be regarded as a feature transformation function with learnable parameters $f(\textbf{X})$, mapping an input feature tensor $\textbf{X}\in\mathbb{R}^{C\times H\times W }$ with spatial size $H\times W$ and $C$ channels to binarized feature tensor $\textbf{X}_{binary}\in\mathbb{R}^{C\times H\times W}$ with more informative features. The function $f(\textbf{X})$ aggregates information from the global receptive field of the network and then computes the thresholds $\alpha^{1:C}$ for each channel in feature map $\textbf{X}$. The calculation of $c$-th element of $\alpha$ can be written as:}

\begin{equation}
    \alpha_{c}=f(\textbf{X}_{c})=\textbf{W}_{2}(\textbf{W}_{1}(\frac{1}{H\times W}\sum_{i=1}^{H}\sum_{j=1}^{W}\textbf{X}_{c}(i,j))) \\
\end{equation}
\textcolor{black}{where $\textbf{W}_{1}\in \mathbb{R}^{\frac{C}{\gamma}\times C}$ and $\textbf{W}_{2}\in \mathbb{R}^{C\times \frac{C}{\gamma}}$ denote two fully-connected (FC) layer. Based the output $\alpha^{1:C}$ from $f(\textbf{X})$, DySign can binarize the  input feature maps $\textbf{X}$ as: } 

\begin{equation}
   \text{DySign}(\textbf{X}_i)=\left\{
    \begin{aligned}
+1 & , & \textbf{X}_{i}>\alpha_{i}, \\
-1 & , & \textbf{X}_{i}\le \alpha_{i}.
\end{aligned}
\right. \\ \alpha_{i}\in \alpha^{1:C} 
\end{equation}
\textcolor{black}{where $\alpha^{i}$ denotes the threshold of \textsl{i}-th channel, which is the \textsl{i}-th element of output vector from $f(\textbf{X})$. The computational unit of DySign can be observed in Fig. \ref{fig:fig3}. The DySign adopts a SEBlock \cite{hu2018squeeze} to learn a set of channel-wise thresholds from the input feature maps for Sign function.  The process is ``$input\to GAP\to FC  layer\to FC layer $''. To avoid over-fit and reduce the extra computational cost, the reduction ratio $\gamma$ is set between two FC layers.} 

\textcolor{black}{For BinaryCCT, the embedding image feature tensor $\textbf{X}\in\mathbb{R}^{C\times H\times W }$ is split into a long sequence $\textbf{X}\in\mathbb{R}^{N\times D }$ with sequence length $N$ and embedding dimension $D$ and then passes through transformer blocks. Compared with BCNNs, BinaryCCT suffers more serious performance degradation since binarization error in each token makes the MSA layer fail to capture distinguishable attention. Thus, DySign can be applied to eliminate this shortcoming. The small difference in the computational unit in DySign between DyBCNN and DyBinaryCCT is that we add the GELU function among two FC layers. In BinaryCCT, we consider generating dynamic thresholds in token-wise. The \textbf{token-wise} threshold $\alpha \in \mathbb{R}^{1\times N}$ is determined by the global information of each token. We apply the DySign in both MSA and FFN layers. Although this function introduces extra FLOPs and memory cost, it can significantly improve the performance of binarized transformers in vision tasks, which will be discussed in Section \ref{section:4}. }  

\textcolor{black}{DySign intrinsically introduces dynamic thresholds conditioned on the input.  This hyperfunction maps the global information of input $\textbf{X}$ to its binarization distribution with more informative features. Based on the simple computational unit in DySign, channel dependencies are explicitly modeled to assign appropriate thresholds for each channel, which can enhance the representation learning of BNNs. In the experiment section, we will show that dynamic learning distribution is an effective and straightforward way to boost the performance of BNNs.}

\subsection{The Architecture of DyBCNN}

Except for the DySign, we also handle the RPReLU function \cite{reactnet2020} in the same way. The shift parameters $\gamma^{1:C}$ and $\zeta^{1:C}$ are learnable based on input feature maps (DyPReLU). The hyper function $f(\cdot)$ generates the shift parameters for each channel in input feature maps. This process can dynamically shift and reshape feature distribution, which is an effective and simple way to increase model capacity. 

Essentially, the DySign can learn the most suitable thresholds $\alpha^{1:C}$ to binarize the input feature map. The threshold parameters can be dynamically adjusted for different input feature maps, which can effectively limit the feature information loss after binarization. For DyPReLU, the $\gamma^{1:C}$ and $\zeta^{1:C}$ can be easily understood as these parameters are dynamically generated to obtain better output distribution. By introducing these functions, the aforementioned problem risen by static parameters can be eliminated. The BCNNs can retain more object information and learn more meaningful features. In the experiment section, we will show that adaptive thresholds learning is an effective and straightforward way to boost the performance of BCNNs.

For model architecture, The ResNet18 \cite{he2016deep} and MobileNetV1 \cite{howard2017mobilenets} are built as the backbones following ReActNet \cite{reactnet2020}. In the ReActNet (Shown as in Fig. \ref{fig:fig2}), the 3$\times$3 depthwise and 1$\times$1 pointwise convolution are replaced by standard 3$\times$3 and 1$\times$1 convolution. The duplication and concatenation operations are designed for addressing the channel number difference. In our case, we simply replace the RSign and RPReLU functions in ReActNet with our DySign and DyPReLU. The ReActNet can be regarded as the baseline in our experiment.

\subsection{The Architecture of DyBinaryCCT}

\begin{figure}[!h]
\centering
\includegraphics[width=8cm]{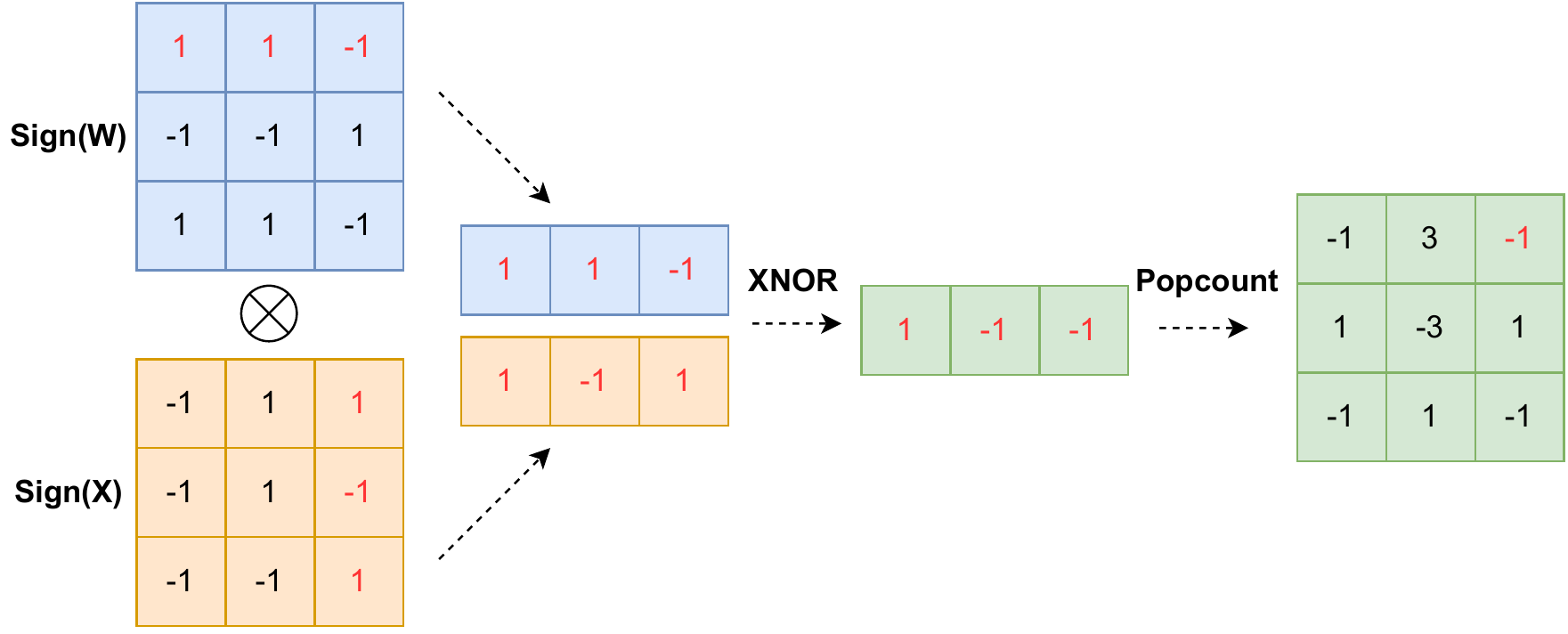}
\caption{Basic XNOR and Popcount operation in the binary linear layer.}
\label{fig:fig5}
\end{figure}

We start with introducing the basic components of BinaryCCT. The tokenization layer and classification head maintain full precision. In each fully binarized transformer block, the matrix multiplication with bitwise XNOR and Popcount operation in the binarized linear layer is presented in Fig. \ref{fig:fig5}.  

For the weights in all binarized linear layers, the scale factor is applied to minimize quantization errors and zero-mean is utilized for retaining representation information \cite{bibert2022}. The activation is only utilized Sign Function to binarize. The complete computational process can be expressed as: 


\begin{equation}
\begin{split}
\text{BiLinear}(X)=\alpha _{w}(\text{Sign}(X)\otimes \text{Sign}(W-u)) \\
\end{split}
\end{equation}
where X and W denote full-precision activations and weights, $u$ is the mean value of weight $W$, $\alpha _{w} = \frac{1}{n}\left \| W \right \|_{\ell1},$ denotes the scaling factor for weight. 

In the BinaryCCT, the input data first passes through a full-precision convolutional block to generate tokens before being fed into multiply transformer encoder layers. The calculation between queries \textbf{Q}, keys \textbf{K} and values \textbf{V} consumes huge computational cost, which can be mitigated in binarized version. For one MHSA layer, the computation in an attention head can be expressed as:   

\begin{equation}
Q, K, V=\text{BiLinear}_{Q,K,V}(X),
\end{equation}
where $\text{BiLinear}_{Q,K,V}(\cdot)$ denotes three binarized linear layers for Q, K, V, respectively. Then the attention score of an attention head can be computed as:
\begin{equation}
\text{A} = \frac{\text{Sign}(Q)\otimes \text{Sign}(K)^{T}}{\sqrt{D}},
\end{equation}
where the $D$ denotes the dimension of features. However, directly binarizing attention matrix after Softmax layer leads to complete information loss since all the elements can be binarized to +1. To address this problem, we follow BiBert \cite{bibert2022} that applies a learnable matrix to shift the distribution in attention matrix. The computation can be expressed as:

\begin{equation}
\text{Attention}(Q, K, V )=\text{Sign}(\text{Softmax}(A)-s)\otimes \text{Sign}(V)
\end{equation}
where the $s$ denotes the learnable matrix. For the rest in MHSA layer and FFN layer in BinaryCCT, we follow the architecture of \cite{cct2021} and replace all full-precision linear layers with binarized linear layers. The DyBinaryCCT is built based on BinaryCCT by replacing all Sign function in MHSA and FFN layers with DySign (See Fig. \ref{fig:fig2}).

\begin{table*}[ht]
\centering
\caption{Compare of the top-1 accuracy with SOTA methods.}
\label{Tab:1}
\begin{tabular}{cccccc}
\hline
Binary Method      & W/A & BOPs($\times10^9$) & FLOPs($\times10^8$) & OPs($\times10^8$) & Acc Top-1 (\%) \\ \hline
BNN\cite{bnn2016}                & 1/1 & 1.70       & 1.20        & 1.47      & 42.2           \\
ABC-Net\cite{lin2017towards}             & 1/1 & $-$          & $-$           & $-$         & 42.7           \\
XNOR-Net\cite{xnor2016}           & 1/1 & 1.70       & 1.41        & 1.67      & 51.2           \\
DoReFa\cite{zhou2016dorefa}             & 1/2 & $-$          & $-$           & $-$         & 53.4           \\
Bi-RealNet-18\cite{birealnet2018}      & 1/1 & 1.68       & 1.39        & 1.63      & 56.4           \\
XNOR++ \cite{bulat2019xnor}             & 1/1 & $-$          & $-$           & $-$         & 57.1           \\
IR-Net \cite{qin2020forward}            & 1/1 & $-$         & $-$           & $-$         & 58.1           \\
BONN\cite{gu2019bayesian}                & 1/1 & $-$          & $-$           & $-$         & 59.3           \\
NoisySupervision\cite{han2020training}  & 1/1 & $-$          & $-$           & $-$         & 59.4           \\
Bi-RealNet-34\cite{birealnet2018}      & 1/1 & 3.53       & 1.39        & 1.93      & 62.2           \\
Real-to-Binary Net\cite{real2020} & 1/1 & 1.68       & 1.56        & 1.83      & 65.4           \\
\textcolor{blue}{ReActNet-ResNet18}\cite{reactnet2020}  & 1/1 & 1.68       & 1.39        & 1.63      & 65.9           \\
\textcolor{red}{ReActNet}\cite{reactnet2020}           & 1/1 & 4.82       & 0.22        & 0.97      & 69.4           \\                       AdamBNN\cite{liu2021adam} & 1/1  & 4.82       & 0.22        & 0.97 & 70.5  \\\hline
\textcolor{blue}{DyBCNN-ResNet18(ours)}     & 1/1 & 1.68       & 1.43           & 1.98         & \textcolor{blue}{67.4$(\uparrow1.5)$}           \\
\textcolor{red}{DyBCNN(ours)}              & 1/1 & 4.82       & 0.24           & 0.99         & \textcolor{red}{71.2$(\uparrow1.8)$}            \\ \hline
\end{tabular}
\begin{tablenotes}
     \item[1] The W/A denote the number of bits in weight and activation quantization. The \textcolor{blue}{blue} font denotes the comparing result with DyBCNN and ReActNet based on ResNet18. The \textcolor{red}{red} font denotes the comparing result based on MobileNetV1.
   \end{tablenotes}
\end{table*}

\subsection{Computational complexity analysis}
Following the calculation method in \cite{birealnet2018,reactnet2020}, we calculate total operations (OPs), which consists of binary operations (BOPs) and floating point operations (FLOPs). The OPs can be obtained as:
\begin{equation}
    OPs=\frac{BOPs}{64}+FLOPs \\
\end{equation}
Taking DyBCNN as an example, we do not introduce extra binary convolutional operations. Thus, the BOPs are the same as ReActNet. The increased computational consumption mainly comes from floating-point operations in DySign and DyPReLU, including one global average pooling layer and two fully-connected layers. For reducing the introduced computational cost, we set the reduction ratio between two fully-connected layers as 16, which can limit the introduced model parameters and float operations. We denote the size of the input feature map as $C\times H \times W$. The FLOPs for each RSign and RPReLU increase $C+\frac{C^2}{8}$ and $2\times(C+\frac{C^2}{8})$. The extra computational cost is only a small fraction of the total cost,  which is above 2\% for DyBCNN and DyBinaryCCT.

\section{Experiments}
\label{Sec:4}
\subsection{Datasets and Implementation details}
\label{cnn experiment}
\textcolor{black}{To evaluate the performance of our proposed method, we conduct experiments on the image classification task and evaluate three datasets: CIFAR10, CIFAR100, and ImageNet dataset \cite{imagenet2015}. The CIFAR-10 dataset consists of 60000 32x32 images, 50000 images for training, and 10000 for testing. The ILSVRC12 ImageNet classification dataset contains 1.2 million training images and 50,000 validation images across 1000 classes, which is more challenging than small datasets. In this section, we first introduce the training  details for DyBCNN and DyBinaryCCT. Then we provide the experimental analysis and performance of the two models separately.}

\textbf{DyBCNN implementation details.} For DyBCNN, we evaluated our method on the ImageNet dataset and compared it with recent state-of-the-art methods. The training strategy utilizes the two-step training strategy described in \cite{real2020}. In the first step, the network is trained from scratch with binary activations and weights. In the second step, the network inherits the weight from the first step and is trained with binary activations and weights. For both steps, we follow the training scheme in \cite{reactnet2020}. We use Adam optimizer and a linear learning rate decay to optimize the model. The initial learning rate is 5e-4, batchsize is set to 256, and $\gamma$ in DySign is set to 16. Besides, we also train a quick version of DyBCNN with the one-step training strategy. 

\textbf{DyBinaryCCT implementation details.} We conducted extensive experiments on three datasets: CIFAR10, CIFAR100, and ImageNet. The training hyper-parameters follow the CCT \cite{cct2021}. All reported accuracy is best out of 4 runs. For CIFAR10 and CIFAR100, the initial learning rate is set to 1e-3 with cosine annealing decay strategy, and weight decay is set to 1e-4. The batchsize is set to 128. The training stage is run for 300 epochs on CIFAR10 and CIFAR100. AutoAugment \cite{aug2019} is used for CIFAR-10 and CIFAR-100. For the ImageNet dataset, the training stage is run for 300 epochs with a batchsize of 64. The initial learning rate is set to 5e-4 with a cosine annealing decay strategy. The weight decay is set to 1e-4. The model is warmed up for 25 epochs. The $\gamma$ in DySign is set to 4.

\subsection{Experiment on DyBCNN}
We compare the DyBCNN with other state-of-the-art binarization methods in Table \ref{Tab:1}. Compared with ReActNet and ReActNet-ResNet18, the DyBCNN and DyBCNN\_ResNet18 achieve 1.8\% and 1.5\% increase, respectively. The superior improvement further proves the importance of activation distribution for BNNs. We also provide the test accuracy of each training step, which can be observed in Tabel \ref{tab:two}. Due to the introduced global average pooling layer and fully-connected layers, the OPs increase slightly. For DyBCNN, Extra OPs from DySign only account for 2\% of the total computational cost, which is acceptable. With the limited computational cost increased, the DyBCNN can outperform previous methods by a large margin, which illustrates the effectiveness of dynamic learnable shift parameters in BCNNs. Besides, another advantage of our proposed method is that it is straightforward and can be easily transferred to any other binarized architecture.

\begin{table}[ht]
\centering
\caption{The accuarcy of DyBCNN in each step}
\label{tab:two}
\begin{tabular}{lcc}
\hline
                & Step 1 & Step 2 \\ \hline
DyBCNN-ResNet18  & 68.4\% & 67.4\% \\
DyBCNN-MobileNet & 72.5\% & 71.2\% \\ \hline
\end{tabular}
\end{table}

To further analyze our proposed method, we conduct following ablation studies for DyBCNN.
\subsubsection{One-step Training}
\label{sec:onestep}
\begin{figure}[!ht]
    \centering
    \includegraphics[width=0.8\linewidth]{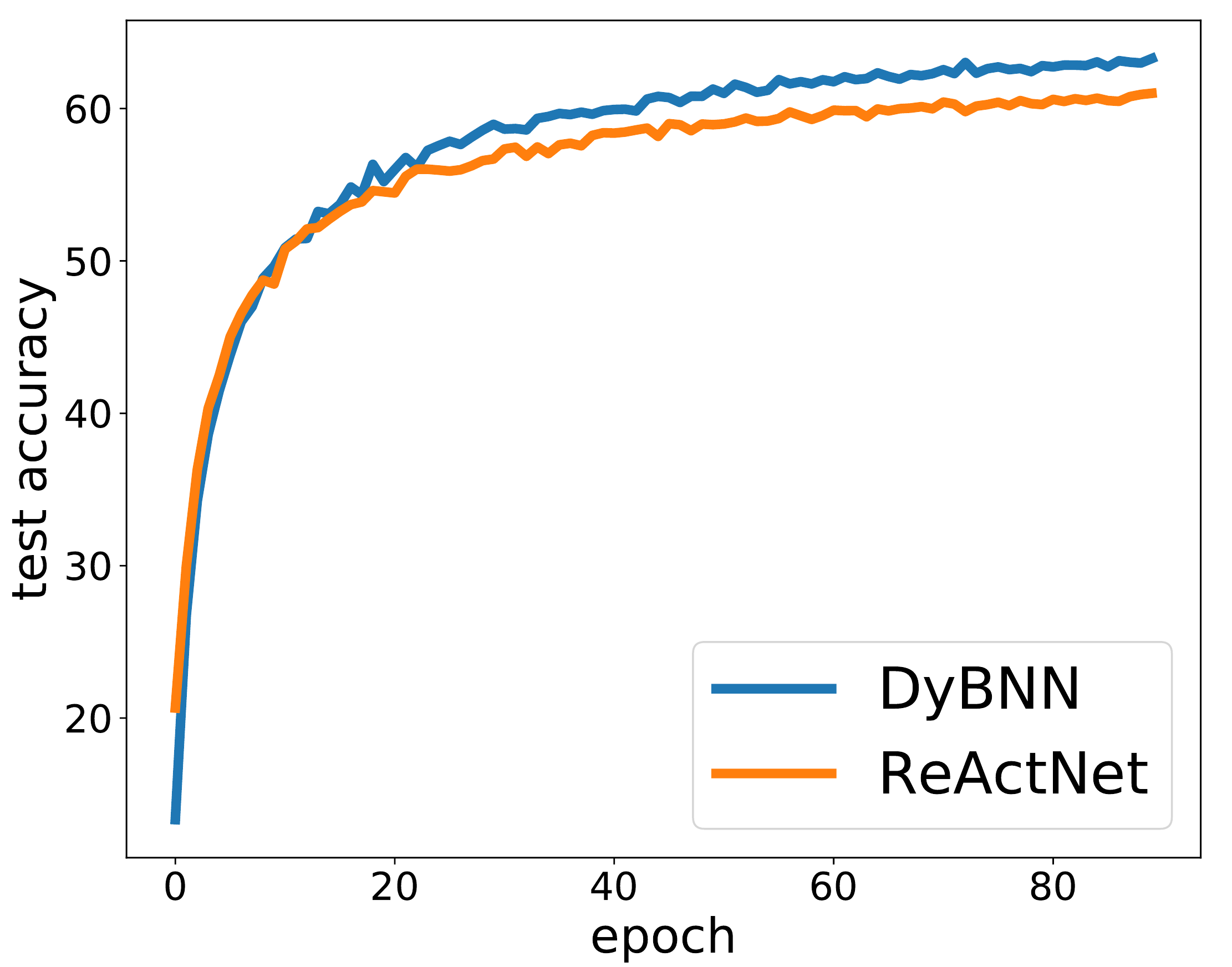}
    \caption{The comparison of accuracy curve with the DyBCNN and ReActNet on ImageNet valid set.}
    \label{fig:Fig6}
\end{figure}

The DyBCNN utilizes the two-step training strategy with 512 epochs in total, which is time-consuming. To simplify the training process, we also evaluate the quick version of DyBCNN in one-step training. The result can be observed in Tabel \ref{Tab:2}. We evaluate DyBCNN under the following strategies: 1) distillation with distribution loss in \cite{reactnet2020}; 2) no distillation with CrossEntropy loss. The training epoch is set to 90. Following other training details described at the beginning of Section \ref{cnn experiment}, DyBCNN achieved 2.3\% and 2.0\% higher than ReActNet under the two training strategies mentioned above, which illustrated that DyBCNN is an effective and straightforward way to boost the performance of BNNs. We also observe that DyBCNN achieves higher accuracy and faster convergence (See Fig \ref{fig:Fig6}). In this paper, The calculation of FLOPs contains BN, PReLU layers, so our reported OPs is higher than $0.87\times10^8$ in \cite{reactnet2020}.

\begin{table}[htbp]
\centering
\caption{The test result of a quick version for DyBCNN.}
\label{Tab:2}
\begin{tabular}{cccc}
\hline
Network  & distillation & OPs                       & Acc Top-1(\%) \\ \hline
ReActNet &              & \multirow{2}{*}{$0.97\times 10^8$} &  61.0             \\
ReActNet &    \Checkmark          &                           & 64.3              \\ \hline
DyBCNN    &              & \multirow{2}{*}{$0.99\times 10^8$} & \textbf{63.3}          \\
DyBCNN    &    \Checkmark         &                           & \textbf{66.3}            \\ \hline
\end{tabular}
\begin{tablenotes}
     \item[1] The \Checkmark denotes the model is trained with distributional loss and the real-valued ResNet34 is set as teacher model. No \Checkmark denotes the model is directly trained with Cross Entropy loss.
   \end{tablenotes}
\end{table}

\subsubsection{Impacts of DySign on DyBCNN}
\label{sec:impact}

 \begin{figure*}[ht]
    \centering
    \includegraphics[width=0.8\linewidth]{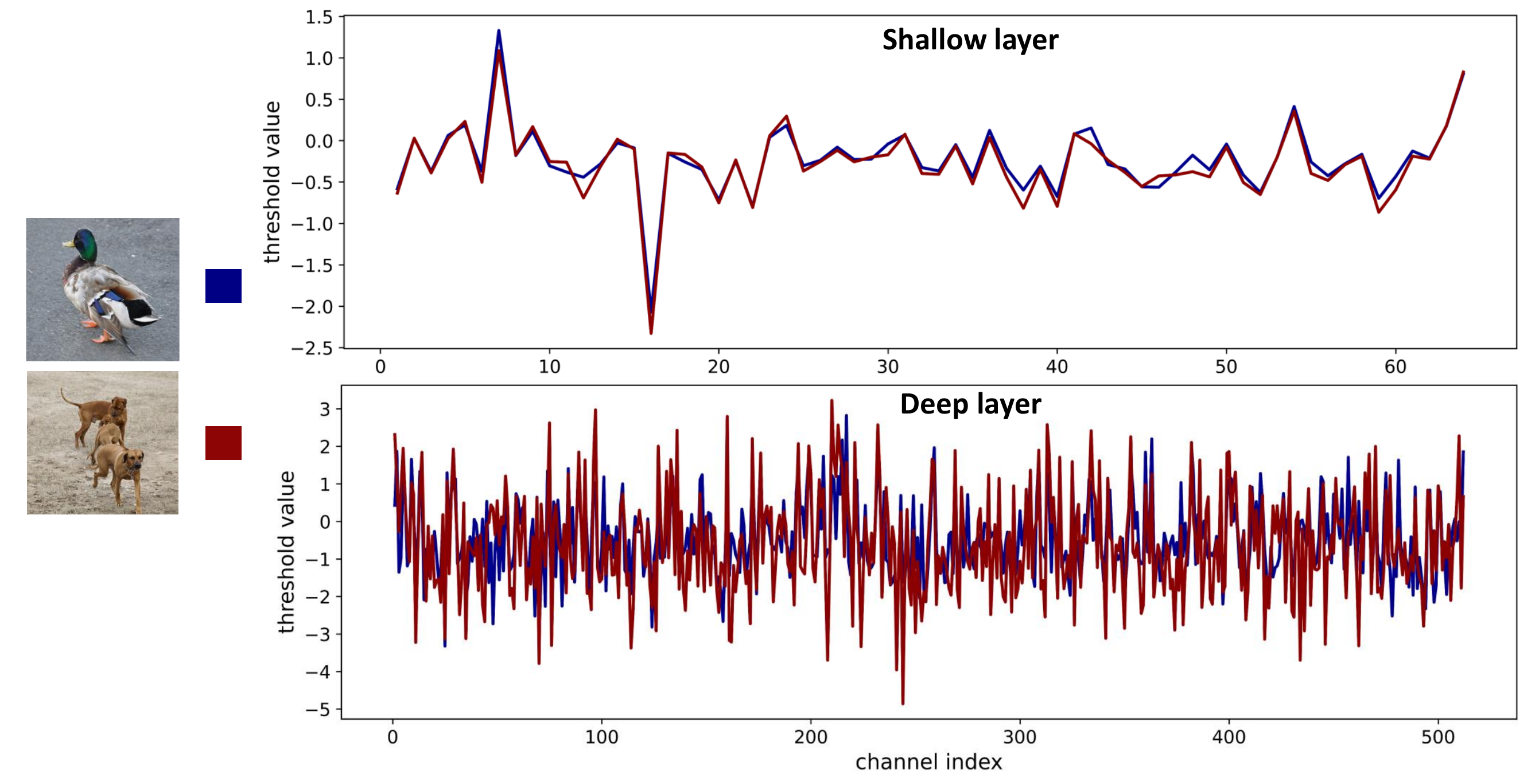}
    \caption{The learning thresholds of DySign for \textcolor{blue}{\textsl{duck}} and \textcolor{red}{\textsl{dog}} samples in shallow and deep layers. The similar patterns are learned in the shallow layers, and diverse thresholds are generated for processing \textcolor{blue}{\textsl{duck}} and \textcolor{red}{\textsl{dog}} features in deep layers.}
    \label{fig:multi}
\end{figure*}

In this section, we analyze the individual effect of DySign. We conduct the experiment in the two-step training strategy following Section \ref{cnn experiment}. The experimental result is shown in Table \ref{Tab:3}. DyBCNN achieves 70.1\% top-1 accuracy on the ImageNet without DyPReLU. Compared with ReActNet, the accuracy increases by 0.7\%, illustrating the effectiveness of our proposed DySign. DySign promotes model performance by reducing feature information loss. Furthermore, DyPReLU increases model expression ability by reshaping the activation distribution of feature maps, which further improves model accuracy. 

\begin{table}[ht]
\centering
\caption{The impact of \textsl{DySign} and \textsl{DyPReLU}.}
\label{Tab:3}
\begin{tabular}{cccc}
\hline
Network & DySign & DyPReLU & Acc Top-1(\%) \\ \hline
ReActNet  &                 &         &    69.4         \\
DyBCNN   &   \Checkmark     &         &    70.1           \\
DyBCNN   & \Checkmark     &    \Checkmark     &   71.2            \\\hline
\end{tabular}
\end{table}

\textcolor{black}{We also visualize the thresholds learned by DySign when processing different samples. As shown in Fig. \ref{fig:multi}, the diverse thresholds can be applied to binarize features when processing the sample \textsl{dog} and \textsl{duck}. We observe that the shallower layer shares  similar learned patterns since they tend to catch the basic and less abstract image features, while the deeper layer diverges different images into different thresholds. The question is raised whether a similar result would be achieved if DySign was only added to deeper layers. Thus, We design an ablation study to verify the effectiveness of our DySign in each layer. We utilize ReActNet based on ResNet18 and conduct experiments on CIFAR100. ResNet18 can be divided into four layers according to the number of channels. We replace RSign with DySign from the last layer forward. The result can be observed in Tabel \ref{Tab:each}. The performance of the model improves as the number of layers incorporated with DySign increases. Compared with adding DySign only to the deepest layer, the accuracy of full use can be improved by 1.3\%. Thus, this flexible method of handling different sample features is helpful for the representation learning of BNNs from shallow to deep layers.} .
\begin{table}[ht]
\centering
\caption{The impact of \textsl{DySign} on each layer.}
\label{Tab:each}
\begin{tabular}{cccccc}
\hline
                        & Layer1 & Layer2 & Layer3 & Layer4 & Accuracy \\ \hline
\multirow{5}{*}{ReActNet-ResNet18} &\XSolidBrush        & \XSolidBrush        &  \XSolidBrush       &  \XSolidBrush       & 66.7\%   \\
                        &  \XSolidBrush       &   \XSolidBrush      &   \XSolidBrush      & \Checkmark      & 68.5\%   \\
                        & \XSolidBrush        &  \XSolidBrush       & \Checkmark      & \Checkmark      & 68.8\%   \\
                        &   \XSolidBrush      & \Checkmark      & \Checkmark      & \Checkmark      & 69.1\%   \\
                        & \Checkmark      & \Checkmark      & \Checkmark      & \Checkmark      & 69.6\%   \\ \hline
\end{tabular}
\begin{tablenotes}
     \item[1] The \Checkmark denotes DySign applied in this layer and \XSolidBrush denotes RSign.
   \end{tablenotes}
\end{table}

\begin{figure}[!ht]
    \centering
    \includegraphics[width=\linewidth]{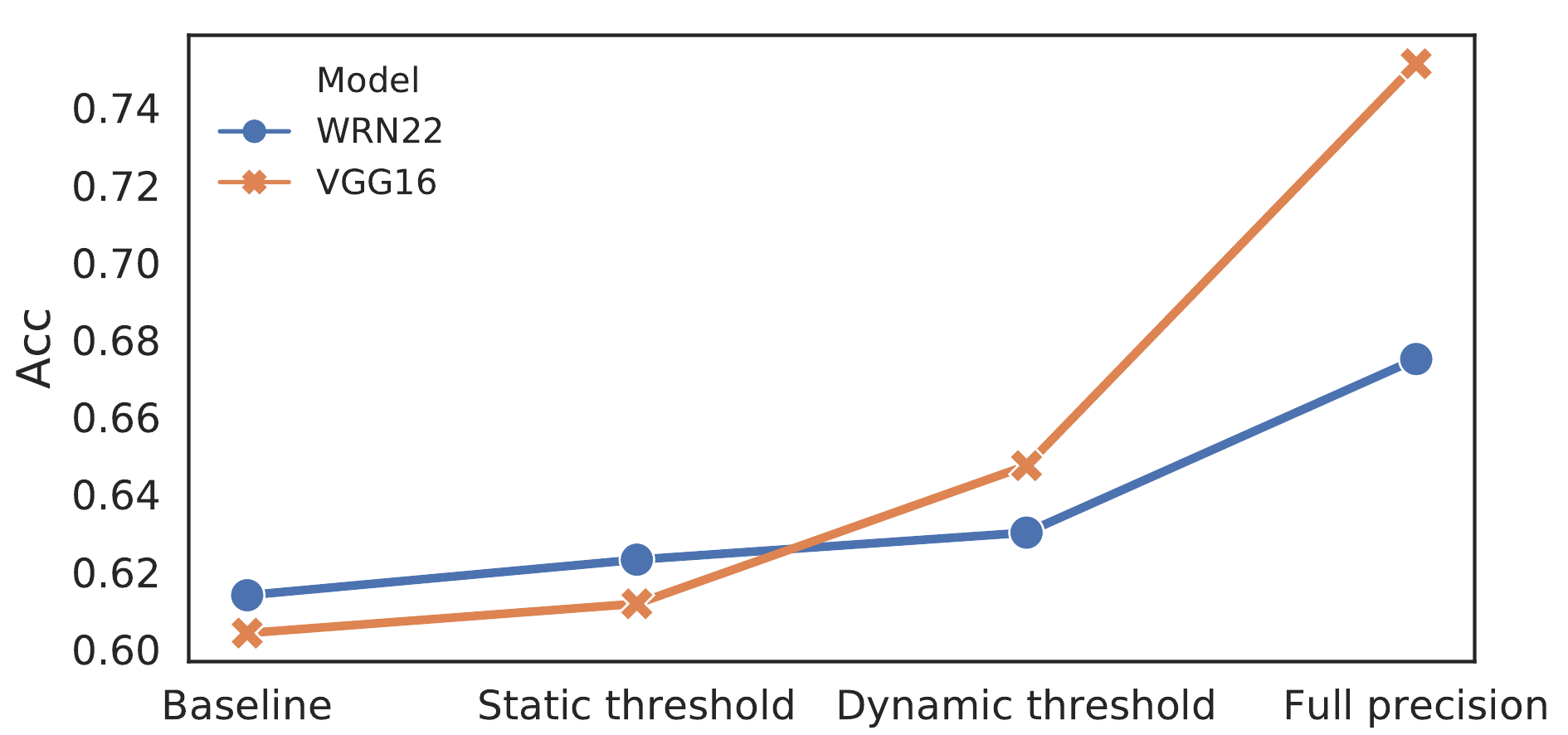}
    \caption{The comparison of VGG16 and WRN22 on CIFAR100 under RSign and DySign. The \textcolor{blue}{blue} line denotes results of WRN22. The \textcolor{orange}{orange} line denotes results of VGG16.}
    \label{fig:Fig7}
\end{figure}

To further evaluate the transfer ability of DySign, we apply our method to two different architectures: WRN22 \cite{zagoruyko2016wide} and VGG16 \cite{simonyan2014very}. WRN is a variant on ResNets that introduces a new depth factor \textsl{k} to adjust the feature map depth expansion through 3 stages, where the spatial dimension of the features remains the same. VGG is a standard deep CNN architecture with multiple layers. In our experiment, we set the \textsl{k} as 2 in WRN22. Thus, the kernel stages for WRN22 and VGG16 are \textsl{32-32-64-128} and \textsl{64-128-256-512-512}, respectively. The DySign is added before each binarized convolutional kernel. We evaluate our method on CIFAR100 shown in Fig. \ref{fig:Fig7}. Compared with the baseline, DySign can improve performance by 1.62\% and 4.33\% for binarized WRN22 and VGG16, respectively. The result suggests that our method can flexibly transfer to other binarized CNN architectures and enhance performance.    

\subsection{Experiments on DyBinaryCCT}
\label{section:4}
In this section, we experimentally analyze the BinaryCCT on both small and large scale datasets (CIFAR10, CIFAR100, and ImageNet datasets). We first compare the accuracy degradation between two different transformer architectures: Vision transformer and convolutional transformer. Then we report the accuracy of our models and implementation details. At the end of this section, we provide the ablation studies to analyze the impact of the introduced DySign.

\subsubsection{ViT or CCT under binarized step}

In this part, we experimentally analyzed two architectures (\textsl{vision transformer and convolutional transformer}) to verify which one is suitable for binarized situations. For the vision transformer, we adopted binarized ViT-Lite introduced from \cite{cct2021}. The ViT-Lite variants with their patch size. For instance, ViT-Lite-7/4 denotes a ViT-Lite with a 7-layer transformer encoder with $4\times4$ patches. For the convolutional transformer, we selected our baseline, binarized compact convolutional transformer (CCT). The CCT is specified by the number of layers, as well as the kernel size for convolutional layers. For instance, CCT-6/$3\times1$ denotes the model has 6 transformer encoder layers and 1 convolutional block with $3\times3$. We binarized the ViT-Lite-7/8, ViT-Lite-7/4, ViT-Lite-6/8, ViT-Lite-6/4, CCT-6/$3\times1$, and CCT-7/$3\times1$.   

\begin{figure}[!t]
\centering
\includegraphics[width=\linewidth]{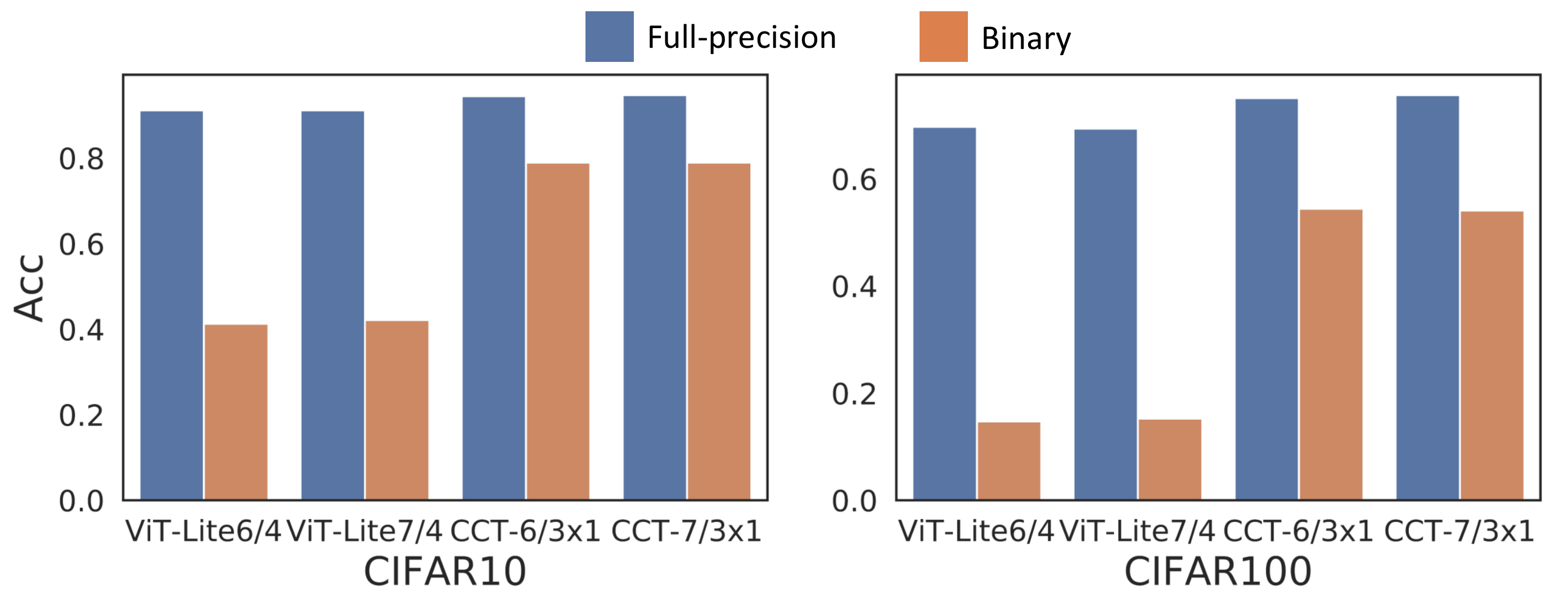}
\caption{\textcolor{black}{The comparison of accuracy degradation under binary operation between ViT architecture and CCT architecture. The accuracy is evaluated on CIFAR10 and CIFAR100. Compared to binarized CCT, binarized ViT suffers a more serious performance degradation.}}
\label{fig:fig8}
\end{figure}

We evaluate these different models on CIFAR10. The comparison of performance degradation between the two architectures can be observed in Fig. \ref{fig:fig8}. The binary operation can significantly influence the module to distinguish attention weights between patches, which suppresses the expression ability of the transformer. The binarized ViT-Lite models suffer catastrophic accuracy degradation (nearly 50\% on CIFAR10 and CIFAR100). For binarized CCT, the accuracy drop is less severe than that of binarized ViT-Lite. This experiment demonstrates that early convolution can effectively boost the performance of binarized vision transformer architecture. For convolutional transformers, setting early visual processing of ViT as convolutions can introduce the inductive bias of CNN, which can enhance the generalization ability of ViT and improve the representation learning ability of transformer blocks. In the binarized setting, a convolution based patching method in the tokenization layer can preserve local relationship information between image tokens. Due to this property, BinaryCCT can remain more valuable information than binarized ViT during binarization. What's more, the low capacity of a binarized linear layer is unable to directly process complicated image information. As a result, the accuracy of BinaryCCT exceeds BiViT-Lite by a large margin, revealing its superiority in the binarized transformer-based module. Thus, we select the convolutional transformer as the binarized module.

\begin{figure}[!t]
\centering
\includegraphics[width=\linewidth]{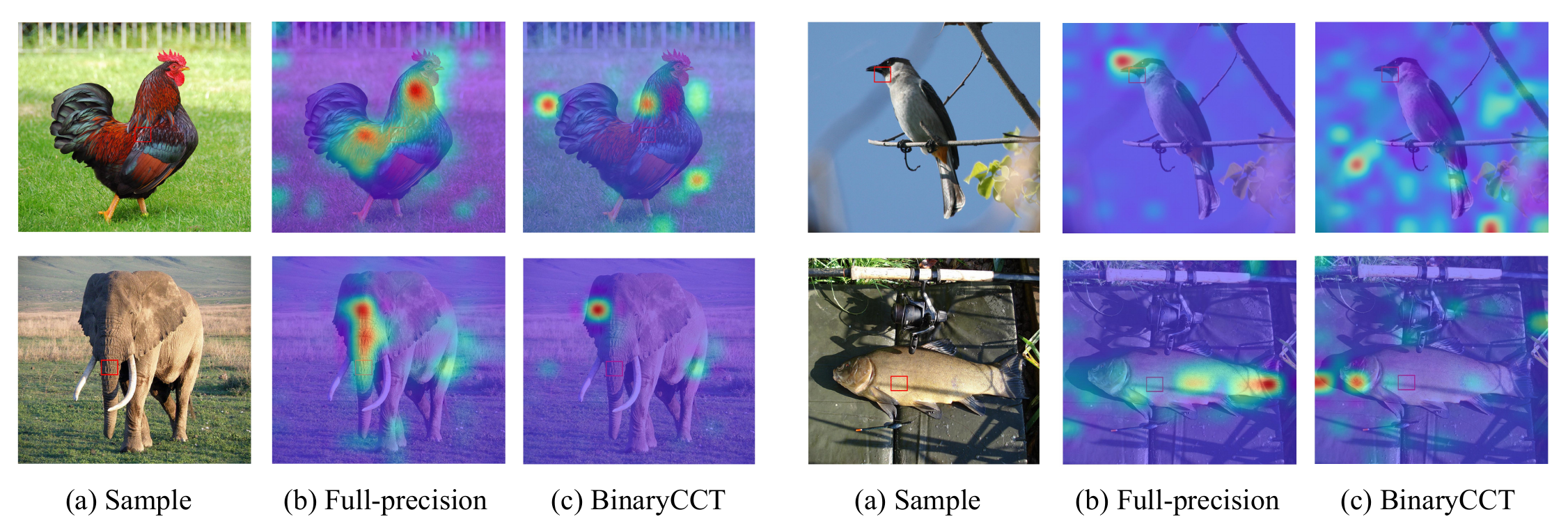}
\caption{The visualization of attention map for (b) full-precision CCT and (c) BinaryCCT. The sample is selected from ImageNet dataset. The sever information degradation leads to the indistinguishable attention.}
\label{fig:fig9}
\end{figure}

\subsubsection{Results on Image Classification}


\begin{table}[ht]
\centering
\caption{The evaluation results on CIFAR10 and CIFAR100.}
\begin{tabular}{lccccc}
\hline
               & CIFAR10 & CIFAR100 & BOPs &FLOPs  & OPs \\ \hline
CCT\_6         & 95.05\%        & 76.95\%         & - & 1.01G      &  1.01G     \\
BinaryCCT\_6   & 78.61\%        & 54.34\%          &1.01G & 7.23M           & 22.96M      \\
DyBinaryCCT\_6 & 84.94\%        & 61.60\%         &1.01G  & 11.16M      & 26.89M      \\\hline
CCT\_7         & 94.76\%        & 76.75\%         & -   & 1.18G   & 1.18G      \\
BinaryCCT\_7   & 78.97\%        &  54.11\%        & 1.17G   & 7.23M       & 25.59M      \\
DyBinaryCCT\_7 & 84.98\%        &  62.10\%        & 1.17G   & 11.81M   & 30.17M      \\\hline
\end{tabular}
\label{tab:tab1}
\end{table}
In the part, we discuss the performance of BinaryCCT on the image classification task. The experiments are conducted on CIFAR10, CIFAR100, and ImageNet. The   
\begin{figure}[ht]
\centering
\includegraphics[width=\linewidth]{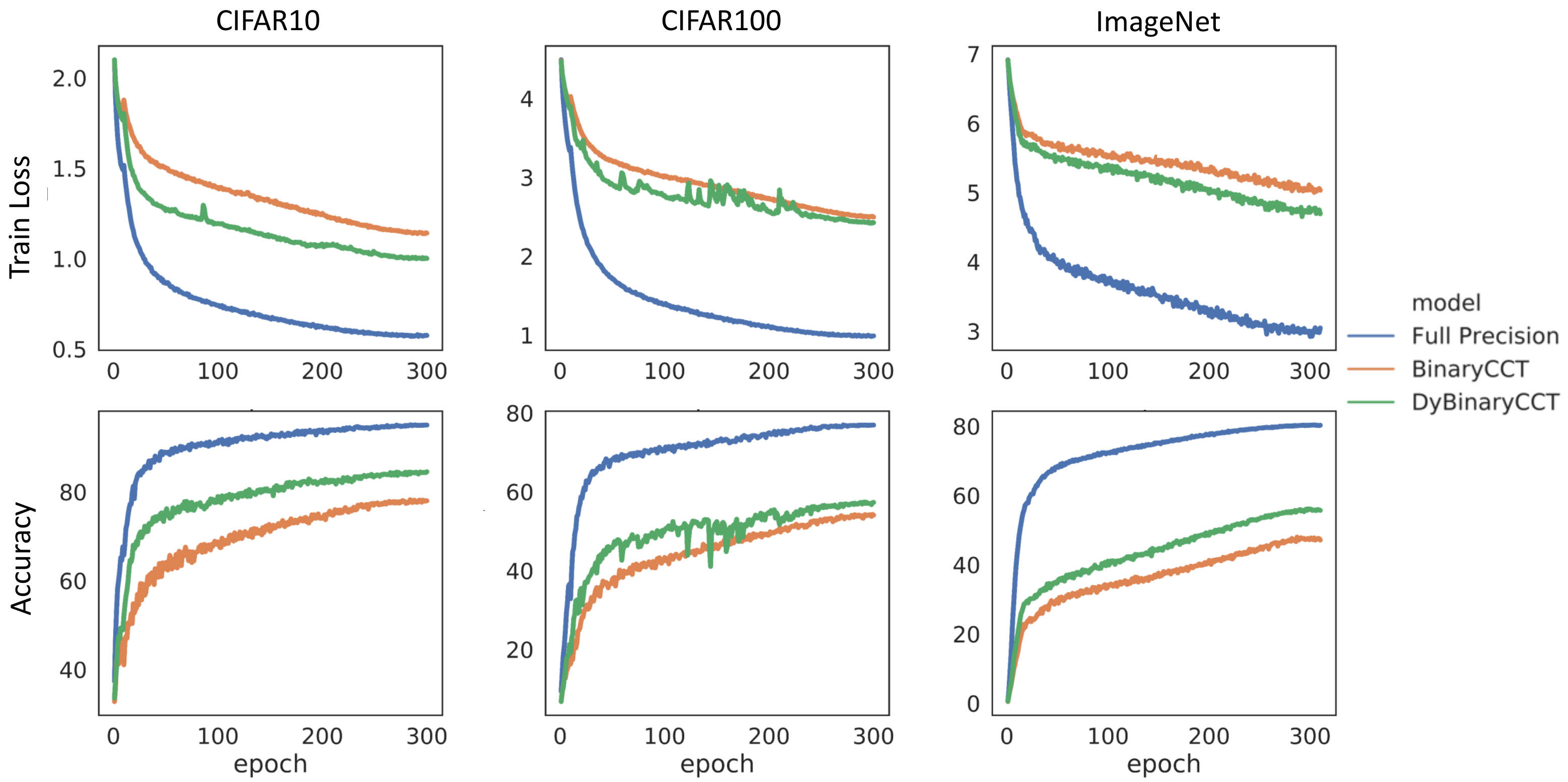}
\caption{Comparisons of training loss and evaluation accuracy of CCT, BinaryCCT, and DyBinaryCCT on CIFAR10, CIFAR100, and ImageNet.}
\label{fig:fig10}
\end{figure}
\begin{table*}[ht]
\centering
\caption{Comparison of different depth and width in BinaryCCT.}
\begin{tabular}{lccccccc}
\hline
\multicolumn{1}{c}{} & \multirow{2}{*}{\textbf{Number of transformer block}} & \multirow{2}{*}{\textbf{embed dim}} & \multicolumn{2}{c}{\textbf{full-precision}} & \multicolumn{2}{c}{\textbf{binarized}} & \multirow{2}{*}{\textbf{OPs}} \\
\multicolumn{1}{c}{} &                                              &                            & top-1            & top-5           & top-1         & top-5         &                        \\ \hline
CCT-7/7$\times$2                & 7                                            & 256                        & 70.17\%          & 90.06\%         & 35.65\%       & 60.11\%       & \textcolor{red}{1.80G}/\textcolor{blue}{0.95G}                        \\ 
CCT-7/7$\times$2               & 7                                            & 384                        &77.62\%                  &93.83\%                 & 41.62\%       & 66.11\%       & \textcolor{red}{2.78G}/\textcolor{blue}{0.97G}                       \\ 
CCT-7/7$\times$2               & 7                                            & 512                        &80.35\%                  &95.03\%                 &47.72\%               &72.02\%               &\textcolor{red}{4.09G}/\textcolor{blue}{0.99G}                        \\
CCT-14/7$\times$2               & 14                                           & 384                        & 80.83\%          & 95.28\%         & 38.22\%       & 63.89\%       &\textcolor{red}{5.40G}/\textcolor{blue}{1.01G}                        \\ \hline
\end{tabular}
\label{tab:tab5}
\begin{tablenotes}
     \item[1] \qquad\quad\quad\quad\quad\textcolor{red}{red} denotes the FLOPs of full-precision model and \textcolor{blue}{blue} denotes the OPs after binarization.
   \end{tablenotes}
\end{table*}

\textbf{Results on small datasets} Firstly, we build our fully binarized transformer baseline BinaryCCT based on CCT-6/3×1 and CCT-7/3x1, named BinaryCCT\_6 and BinaryCCT\_7. The BinaryCCT When evaluated on BinaryCCT\_6, the accuracy drops 16\% on CIFAR10 and more than 20\% on CIFAR100 (See Tabel \ref{tab:tab1}). This severe performance degradation derives from indistinguishable attention in binarized transformer blocks according to the observation in Fig. \ref{fig:fig9}. The training loss curves (See Fig. \ref{fig:fig10}) also present the sluggish convergence of BinaryCCT. Besides, due to low model capacity, the loss curve is unable to continue to fall especially on the large-scale dataset ImageNet. 

To enhance the performance of BinaryCCT, we apply DySign on the baseline models. The learnable thresholds have been proved as an effective method to enhance BNN accuracy \cite{reactnet2020,dybnn2022}. As shown in Tabel \ref{tab:tab1}, the accuracy of DyBinaryCCT\_6 is improved by nearly 6\% on CIFAR10 and nearly 8\% on CIFAR100. The DySign is a straightforward and effective way to reduce information loss and boost the representational ability of BinaryCCT. The detail analysis can be observed in Section \ref{Sec:4.3}. 

\textbf{Results on large datasets} Except the small dataset, we also explore traits of BinaryCCT on large-scale dataset ImageNet. Due to the limitation of computing resources, we selected CCT-7/7$\times$2 as our baseline model, which consists of 7-layer transformer block and 2-layer convolutional embedding layer. The experiment is shown in Tabel \ref{tab:tab5}. Compared with baseline, BinaryCCT only achieves 35.65\% top-1 accuracy, accuracy loss of approaching 40\%. Based on the observation in Fig. \ref{fig:fig10}, the sluggish rate of descent loss and convergence speed demonstrate that BinaryCCT is unable to distinguish property for complicate object categories due to low-capacity. We attempt to improve the capacity from two perspectives: depth and width. For depth, we increase the transformer blocks, \emph{eg.} 14-layer. For width, we wide the dimension in each transformer block, \emph{eg.} 384 and 512.

In this paper, we utilized CCT-14/7$\times$2, increasing number of transformer block to 14 to evaluate the impact of depth. And we set the embedding dimension as 256, 384, 512 to evaluate the impact of width. The experimental results observed in Tabel \ref{tab:tab5} suggest increasing embed dimension of transformer block is an effective way to boost performance of BinaryCCT. When the dimension is extended from 256 to 384, the accuracy of the model increased by 6.03\%. For adding transformer block, the accuracy of 14-layer BinaryCCT is 3.4\% lower than 7-layer BinaryCCT. Due to the severe loss of semantic information, the model is unable to capture the correct relationship between tokens. Therefore, deepening binarized transformer layers fails to enhance the representation ability of the model and confuses classifier by indistinguishable attention. Besides, the matrix multiplication in transformer blocks decompose to XNOR PopCount operation. Thus, increasing width of BinaryCCT only introduces negligible amount of computational cost. As shown in Table \ref{tab:tab5}, the \textcolor{red}{red} denotes the FLOPs of full-precision model and \textcolor{blue}{blue} denotes the OPs after binarization. When extending embedded dimension from 256 to 512, OPs of BinaryCCT only increases 0.04G while achieving significant performance improvement.

The experiment of DyBinaryCCT is shown in Tabel \ref{tab:tab6}. We set the embedding dimension to 512. With DySign, DyBinaryCCT can achieve 56.09\% top-1 accuracy, nearly 9\% higher than the baseline. The result suggests the importance of retaining token information for the binarized transformer performance. The single-threshold leads to the MHSA layer being unable to capture the relationship between each token since excessive noise and information loss exist in binary tokens. DySign reduces the negative effects of binary operation by adjusting the distribution of tokens so that the attention module can capture the accurate dependencies between tokens. 

For the computational cost, although floating-point operations of the convolutional tokenization layer occupy almost 70\% of the FLOPs and affect the overall calculation consumption, the computational cost of binarized transform blocks is only 2\% of the full-precision version. Therefore, deepening or widening the BinaryCCT to further improve the performance will not introduce too much computation.

\begin{table}[hb]
\centering
\caption{The result of DyBinaryCCT}
\begin{tabular}{lccccc}
\hline
Model       & Top-1 Acc & Top-5 Acc & BOPs & FLOPs & OPs \\ \hline
Full-precision   &80.35\%           &95.03\%           & -  & 4.10G      & 4.10G      \\ 
BinaryCCT   &47.72\%           &72.02\%           & 3.15G   & 0.94G    & 0.99G      \\ 
DyBinaryCCT &56.09\%           &79.14\%           & 3.15G  &  0.95G    & 0.99G     \\ 
 \hline
\end{tabular}
\label{tab:tab6}
\end{table}

\begin{table*}[!hb]
\centering
\caption{The comparison of RSign and DySign}
\begin{tabular}{lcccccc}
\hline
                                & static\_channel & static\_token & dynamic\_channel & dynamic\_token & CIFAR10 & CIFAR100 \\ \hline
\multirow{4}{*}{DyBinaryCCT\_6} & \Checkmark               &               &                  &                &  78.50\%       & 53.59\%         \\
                                &                 & \Checkmark             &                  &                & 80.52\%        & 55.92\%         \\
                                &                 &               & \Checkmark                &                &\textbf{85.10\%}         &60.52\%          \\
                                &                 &               &                  & \Checkmark              &84.94\%         &\textbf{61.60\%}          \\ \hline
\multirow{4}{*}{DyBinaryCCT\_7} & \Checkmark               &               &                  &                &78.67\%         &52.42\%          \\
                                &                 & \Checkmark             &                  &                & 80.65\%        &  56.98\%         \\
                                &                 &              & \Checkmark                &                & \textbf{85.37\%}        & 59.68\%          \\
                                &                 &               &                  & \Checkmark              &   84.98\%      & \textbf{62.10\%}         \\ \hline
\end{tabular}
\label{tab:tab3}
\end{table*}
\subsubsection{The impact of DySign in Transformer block}
\label{Sec:4.3}

In this section, we provided a detail analysis of the DySign. The ablation study is conducted on CIFAR10 and CIFAR100. To demonstrate the superiority of architecture, we also compared with RSign introduced from \cite{reactnet2020}. As shown in Table \ref{tab:tab3}, the learnable thresholds can effectively improve the representation ability of CNN by eliminating the information loss in the binarized operation, especially on CIFAR100. For RSign, accuracy improvement on CIFAR100 is about 2\% compared with the baseline, which is 6\% lower than DySign. The comparison in Tabel \ref{tab:tab3} demonstrates that the static threshold is not sufficient to compensate for the accuracy gap between full-precision model due to the information loss and low capacity of binarized linear layers. The dynamic threshold learning module adaptively generates the thresholds based on input tokens, which effectively reduce information loss and boost the representational ability of binarized transformer block.  

Furthermore, We compare the performance of applying DySign channel-wise and token-wise. Different from token-wise DySign (thresholds $\alpha \in \mathbb{R}^{1\times N}$), the channel-wise thresholds $\alpha \in \mathbb{R}^{1\times D}$ is determined by the average information in each embedding dimension of input tokens. As shown in Tabel \ref{tab:tab3}, the performance of the two methods is almost the same on CIFAR10. When evaluated on CIFAR100, the token-wise dynamic thresholds can significantly enhance the performance of BinaryCCT, which is 2.42\% higher than channel-wise dynamic thresholds on DyBinaryCCT\_7. The experimental result illustrates that token-wise thresholds are suitable for the transformer block to process feature information and retain the representation of each token. Thus, we select token-wise dynamic thresholds in our method. 

\section{Conclusions}
\label{Sec:5}
\textcolor{black}{In this paper, we first propose a simple and effective computational unit to adaptively binarize features in BNNs, named as DySign. DySign explicitly models the channel dependencies to assign appropriate thresholds for each channel, which can effectively reduce the feature information loss and enhance the representation learning of BNNs. Furthermore, we proposed a fully binarized transformer architecture for vision tasks, named BinaryCCT. We evaluate the proposed DySign on BCNN and BinaryCCT, two popular backbones in the visual recognition tasks. The extensive results suggest that our method can significantly improve the performance of BCNN and BinaryCCT with a limited computational cost increase.}

\textbf{Future works.} During our experiment, we observe two points: 1) binarizing $1\times1$ convolution and linear layer results in significant degradation of model accuracy. 2) The knowledge of the full-precision teacher model cannot effectively transfer to the binarized student model. We assume that 1) low-capacity 1-bit $1\times1$ convolution and linear layer fail to process complicated semantic features; 2) the binarized student model is hard to mimic the distribution of a full-precision network. In the future, we will consider improving the performance of binarized vision architecture from these two points.

\bibliographystyle{IEEEtran}
\bibliography{IEEEabrv,binarycct}

\end{document}